\documentclass[12pt,preprint]{aastex}
\usepackage{emulateapj5}
\newcommand{\ms}{\mbox{m\,s$^{-1}$}\,}

\newcommand{\mss}{\mbox{m${^2}$\,s$^{-2}$}\,}

\newcommand{\muher}{\mbox{$\mu$~Her}}

\newcommand{\Dnu}{\Delta \nu}
\newcommand{\half}{{\textstyle\frac{1}{2}}}
\newcommand{\muHz}{\mbox{$\mu$Hz}}
\let\epsilon\varepsilon
\shorttitle{Oscillations in $\mu$ Her}
\shortauthors{Bonanno et al.}
\begin{document}

\title{Detection of solar-like oscillations in the G5 subgiant \muher}

\author{
Alfio  Bonanno,\altaffilmark{1}
Serena Benatti,\altaffilmark{2}
Riccardo Claudi,\altaffilmark{2}
Silvano Desidera,\altaffilmark{2}
Raffaele Gratton,\altaffilmark{2}
Silvio Leccia,\altaffilmark{3}
Lucio Patern\`o \altaffilmark{4,1}
}

\altaffiltext{1}{
INAF, Osservatorio Astrofisico di Catania, Via S. Sofia 78, I-95123 Catania, 
Italy; abo@oact.inaf.it 
}
\altaffiltext{2}{
INAF, Osservatorio Astronomico di Padova, Vicolo Osservatorio 5, I-35122 Padova, Italy 
}
\altaffiltext{3}{
INAF, Osservatorio Astronomico di Capodimonte, Salita Moiariello 16, I-80131 Napoli, 
Italy }
\altaffiltext{4}{
Dipartimento di Fisica ed Astronomia, Universit\`a di Catania, 
Via S. Sofia 78, I-95123 Catania, Italy}
\begin{abstract} 
A clear detection of excess of power, providing a substantial evidence
for solar-like oscillations in the G5 subgiant \muher{}, is presented. 
This star was observed over seven nights with the SARG echelle spectrograph 
operating with the 3.6-m Italian TNG Telescope, using an iodine absorption 
cell as a velocity reference.  A clear excess of power
centered at 1.2\,mHz, with peak amplitudes of about 0.9\,\ms in the amplitude spectrum
is present.  Fitting the asymptotic relation to the power spectrum, 
a mode identification for the $\ell=0,1,2,3$ modes in the frequency range 
$900-1600\; \muHz$ is derived. The most likely value for the large separation turns 
out to be  56.5 \muHz, consistent with theoretical expectations.    
The mean amplitude per mode ($l=0,1$) at peak power results 
to be $0.63 \;\rm m\,s^{-1}$, almost three times larger than the solar one. 
\end{abstract}


\section{Introduction}
The search for solar-like oscillations in stars has experienced a tremendous growth 
in recent years   (See \citealt{B+K2006b},  for a summary).  Most of the results came
from high-precision Doppler measurements using spectrograph such as
CORALIE, HARPS, UCLES, UVES.  
In particular, recent measurements obtained with the SARG spectrograph have led
to the determination of frequencies, mode amplitudes, lifetime and granulation
noise on Procyon A \citep{cla05,sil07}, a star for which the nature of oscillation
spectrum is still debated \citep{natureHD,nature2004}. 

In this paper we report the detection of excess of power, providing evidence for
oscillations, in the G5 subgiant \muher{} (HR~6623, $V=3.417\pm0.014$, G5~IV), a
slightly evolved  solar-type star with mass $1.1\,M_\sun$,  
effective temperature $T_{\rm e}=5596 \pm 80 \; \rm K$,  and $\rm log \; g = 3.93\pm 0.10$
\citep{fu98}.    

\section{Observations and data reduction}
The observations were carried out over seven nights (2006 June 13--19) 
with the  high resolution cross dispersed echelle spectrograph SARG (\citealt{gratton01}) 
mounted on the Italian 3.6m telescope TNG at La Palma observing
station.  Our spectra were obtained at $R=144,000$ in the wavelength range between 462 and 
$792\,\rm nm$, where the calibration iodine absorbing cell covers only the blue part  of the spectrum
(from 462 up to $620\,\rm nm$). Exposure times were typically 60\,s, with a dead-time of 55\,s between
exposures due to the readout time.  The signal-to-noise ratio for most
spectra was in the range from 200 to 400, depending on the seeing and
extinction.  In total, 1106 spectra were collected, with the following
distribution over the seven nights: 27, 106, 183, 186, 180, 226, 198.  The first two
nights were affected by poor weather and technical problems. 

The extraction of radial velocities from the echelle spectra was based on the  
method described by \citet{endl}.  
The observed spectrum was fitted with a reconstructed one, by
using a convolution between the oversampled stellar template, 
the very high resolution iodine cell spectrum and the measured spectrograph instrumental profile. 
Essential to this process are the template spectra  of \muher{} taken
with the iodine cell removed from the beam, and of the
iodine cell itself superimposed on a rapidly rotating B-type star, which 
was the same for all the measurements. 

\begin{figure*}
\epsscale{.8} 
\plotone{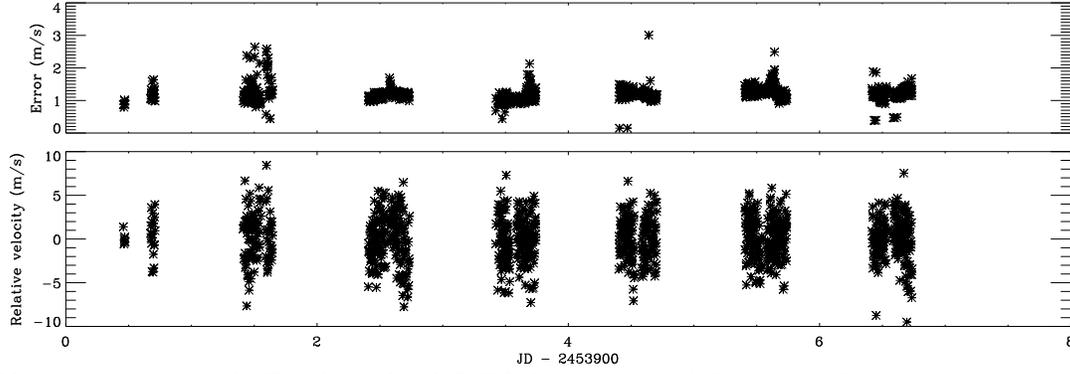}
\caption[]{\label{fig1}Velocity measurements of \muher{}
obtained with  SARG (lower panel) and the corresponding uncertainties
(upper panel).}
\end{figure*}

The resulting velocity measurements of \muher{} are shown in the lower
panel of Fig.~\ref{fig1}.  They have been corrected to the solar
system barycenter and no other correction, such as
decorrelation or high-pass filtering has been applied. 
The rms scatter of these measurements is 2.53\,\ms  
and the uncertainties for the velocity measurements were estimated from residuals
in the radial velocity extraction procedure and are shown in the upper panel of
Figure.~\ref{fig1} where it can be noticed that
most lie in the range 1--2\,\ms, except for the second 
night where technical problems due to the guiding system occurred.  
\section{Time series analysis}\label{sec1}
The amplitude spectrum of the velocity time series was calculated as a
weighted least-squares fit of sinusoids \citep{FJK95,AKN98,BKB2004,KBB2005}, with a weight
being assigned to each point according to its uncertainty estimate obtained from
the radial velocity measurement.
We have then optimized the weight following the approach  discussed in  
\citealt{BBK2004}.  This consists in (i)~cleaning  all power at low
frequencies (below 250\,\muHz) from the time series, as well as all power from oscillations
(800--1800\,\muHz); and (ii)~searching these residuals for points that
deviated from zero by more than it would be expected from their uncertainties.
We found that 20 data points  needed to be significantly down-weighted.  

Fig.~2 shows a pronounced excess of power in the power spectrum (PS) 
around 1.2\,mHz which is the clear signature of
$p$-modes oscillations for a G5 subgiant star.  This feature is apparent in the power spectra of
individual nights, and its frequency and amplitude are in  agreement with 
theoretical expectations, as we shall discuss. 
Moreover, it is clearly disentangled from the low frequency increase 
due to slow instrumental drifts ($1/f$ behavior in the amplitude spectrum) and
the high frequency white noise. In particular, we
find that the mean noise level in the amplitude spectrum in the range 3--5\,mHz is $\sigma = 0.1$\,\ms 
which correspond to a mean noise level in the PS of 0.013\,\mss.  
Since this is based on 1106 measurements, we can deduce 
that the velocity accuracy on the corresponding timescales is 1.82\,\ms.
%
%
\subsection{Large frequency spacing and oscillation frequencies}
The mode frequencies for low-degree, high radial order $p$-mode oscillations 
in Sun-like stars  are reasonably well
approximated by the asymptotic relation (\citealt{tassoul}):
\begin{equation}
  \nu(n,l) = \Dnu{} (n + \half l + \epsilon) - l(l+1) \delta\nu_{02}/6
        \label{uno}
\end{equation}
where $n$ and $l$ are integers which define the radial order and angular
degree of the mode, respectively; $\Dnu{}$ (the so-called large frequency separation)
reflects the average stellar density, $\delta\nu_{02}$ is sensitive to the sound speed
gradient near the core, and $\epsilon$ is a quantity of the order unity 
sensitive to the surface layers.  
Note that $\delta\nu_{02}$ is the so-called small
frequency separation between adjacent modes with $l=0$ and $l=2$.

On attempting to find the peaks in our power spectrum matching the asymptotic
relation, we were severely hampered by the single-site window function.  As
is well known, daily gaps in a time series produce aliases in the power
spectrum at spacings $\pm11.57\,\muHz$ and multiples, which are difficult to disentangle
from the genuine peaks.  Various methods for the search of regular series of peaks
have been discussed in the literature, such as
autocorrelation, comb response and histograms of frequencies.
We used the comb response function method
where a comb response function $CR(\Dnu{})$ is calculated for all sensible
values of $\Dnu{}$ (See \citealt{kjeldsen95} for details).
In particular we used the following function:
\begin{equation}	\label{comb}
CR(\Dnu{})=\prod_{n=1}^{5}\Big [ PS\big (\nu_{\rm max}\pm \frac{2n-1}{2} \Dnu{} \big)
PS\big (\nu_{\rm max} \pm n \Dnu{} \big ) \Big ]^{\frac{1}{2^{n-1}}} \label{cr}
\end{equation}
so that a peak in the $CR$ at a particular value of $\Dnu{}$ indicates the
presence of a regular series of peaks in the PS, centered at $\nu_{\rm max}$
and having a spacing of $\Dnu{}/2$.  
It differs from a correlation function as the product of individuals terms is
considered rather than the sum. Moreover it 
generalizes the  $CR$ defined in \citet{kjeldsen95} including  
(and down-weighting) four outermost  additional frequencies  so that  
it better detects the presence of a regular series of peaks as we checked in  
simulated power spectra. 

To reduce the uncertainties due to the noise, only frequencies  
with amplitude greater than $4\,\sigma$ in the amplitude spectrum,   
in the frequency range $800-1800 \, \mu\rm Hz$,
have been used to compute the $CR$ \label{cr}. 
We then determined the local maxima of the response function $CR(\Dnu{})$  
in the range $40 \leq \Dnu{} \leq 80\; \rm \mu Hz$ and   
the resulting cumulative comb response  function, 
obtained by summing the contributions of all the response functions,  
had the most prominent peak  at $57 \rm \mu Hz$ as  shown in Fig.~3.   
  
The large separation $\Dnu{}$  scales approximately with the square root of
the mean density of the star (\citealt{cox}) and by extrapolating from  the solar case 
($\Dnu{}=134.8 \; \muHz$; see  \citealt{be07}) one obtains 
$\Dnu{} = (M/M_\sun)^{1/2}(R/R_\sun)^{-3/2} \; \Delta\nu_\sun = 59 \,\pm 4\muHz $  
where the error derives from the uncertainties on the radius quoted
in \citet{fu98}. Since  we searched over the entire range
40--80\,\muHz, the agreement between observation and theory is encouraging. 

We tried to extract  the oscillation frequencies directly from the PS 
by using the modified extraction method  described in \citet{kjeldsen95}
and recently used in \cite{sil07}: for each frequency $\nu_{\rm max}$ extracted in the step before, 
a frequency region of $2\,\rm\mu Hz$ width centered
on $\nu_{\rm max}\pm n \; 57\,\rm\mu Hz$, with $n=1,...$ was selected.   
When this region contained a peak with an amplitude greater or equal than
 $3\,\sigma$ in the amplitude spectrum we identified this peak as the second 
component to be cleaned, and recomputed the PS for each $n$. 
The advantage of using this approach is that the extracted frequencies are determined  
except for the same  day-night alias shift for each group of frequencies 
extracted during the procedure.  
In order to identify the individual frequencies of the modes we then constructed  echelle diagrams 
corresponding to values of the large separation around $57\, \rm\mu Hz$, so that we could
easily identify the ridges of the  $\ell =0$ and then those of $\ell =1,2,3$.

It is important to see if  there are additional frequencies, not following the comb-like structure, which where not found with the modified extraction.  We thus analyzed the spectrum 
of the residuals obtained with the modified extraction, 
as shown in the upper panel of Fig.~4, and we found 5 additional frequencies with amplitude 
$ \geq 3\,\sigma$ namely,
$\nu_1 =1133.6 \; \rm \mu Hz$;  
$\nu_2 =1580.5 \; \rm \mu Hz$;  
$\nu_3 =1387.1 \; \rm \mu Hz$;   
$\nu_4 =1172.9 \; \rm \mu Hz$ and  
$\nu_5 =1045.6 \; \rm \mu Hz$. 
These frequencies can be of stellar origin but can also be spurious peaks due to the noise
or due to the non-linearity of the extraction procedure. 
On the other hand it is possible to note that that $\nu_4$ can be  
identified with an $l=3, n=18$ mode and $\nu_5-11.57=1034.03$ can be identified with a 
$l=2, n=16$ mode.
The left panel of Table~1  shows the identified frequencies according with the modified
extraction. It should be noticed that although we define a peak to be significant
if its amplitude is greater or equal than $3\,\sigma$ in the amplitude spectrum, in Table~1 
we have included one frequency with amplitude $2.9 \, \sigma$ as this frequency appears
also with the standard extraction method (see below) and with an higher signal to noise 
ratio. 

The large and small frequency separation and the constant $\epsilon$
have thus been determined by means of a least square best fit with
the asymptotic relation (\ref{uno}) for all the identified modes by means of the modified extraction,
which provided us with:  $\Dnu{} = 56.46 \pm 0.11 \;  \mu \rm Hz$,  
$\delta\nu_0 = 5.22 \pm 0.81 \; \mu \rm Hz$ and $\epsilon=1.46\pm 0.05$.
The expected position of the modes according to this solution is depicted 
with dashed lines in the left panel of Figure~5 while  
Figure~6 shows a close-up of the power spectrum with the identified 
frequencies according with the modified extraction. 

As deviations from Eq.~(1) could be expected for  $\ell \not = 0 $  
in the case of avoided crossings (\citealt{ASW77}),  it is important to 
check that the standard iterative sine-wave fitting leads to a consistent identification. 
We thus extracted iteratively all the frequencies with amplitude greater than $3\, \sigma$
and tried to construct an echelle diagram by searching in the range $50-60 \; \mu \rm Hz$. 
It turned out that only with $\Dnu  \approx 57$ it was possible to clearly see the 
ridge corresponding to $\ell=0$. It was then straightforward to identify the $\ell=1,2,3$ mode,
except for two frequencies, namely 
$\nu_1 =1112.0 \; \rm \mu Hz$,   
$\nu_2 =1596.6 \; \rm \mu Hz$  
which could not be identified. The PS of the 
residuals in shown in the lower panel of Fig.~4
while the resulting echelle diagram is depicted in the right panel of Fig.~5.

A synoptic view of the extracted frequencies with the two methods is then presented in Table~1 
where the frequencies extracted according with the modified extraction (left)
are compared with those obtained with the standard extraction (right).
It is reassuring to notice that 11 frequencies have the same identification in 
both approaches. In particular the identification of the $\ell=0$ 
ridge appears to be quite robust, and indeed one can see in Fig.~5 
that the echelle diagrams obtained
with the modified extraction (left) and the standard extraction (right) 
are essentially the same; this result suggests that we have correctly detected the large separation
in this star. 

In Table~1 the error on frequencies is reported in parentheses. 
In particular we have used the analytical estimates 

\begin{eqnarray}
&&\sigma(a)=\sqrt{\frac{2}{N}} \; \sigma(v)\simeq 0.1 \; {\rm m/s}, \nonumber\\
&&\sigma(\nu) = \sqrt{\frac{6}{N}}\; \frac{1}{\pi T} \;  \frac{\sigma(v)}{a}
\simeq \frac{0.1}{a_{m/s}} \; \mu \rm Hz
\end{eqnarray}

\citep{mgn}
where $N=1106$ is the number of data points, $T=6.25$ is the total duration of the run in
days  and $\sigma(v) = \langle \sigma_v \rangle$ is the mean error of the data.
Note that $\sigma(a)$ represents the noise of the amplitude spectrum after pre-withening 
and $\sigma(a)/a$ is the inverse of the $S/N$ ratio. However 
the errors given by these formulae represent a lower limit to the real
error \citep{mgn}. In particular the actual errors can be several times larger up to a factor 
$\approx 10$ (note that the formal resolution is  $1.8 \rm \; \mu Hz$).
 
The large and small frequency separation and the constant $\epsilon$
have thus been determined by means of a least square best fit with
the asymptotic relation (\ref{uno}) for all the identified modes,
which provided us with:  $\Dnu{} = 56.50 \pm 0.07 \;  \mu \rm Hz$,  
$\delta\nu_0 = 5.03 \pm 0.94 \; \mu \rm Hz$ and $\epsilon=1.44\pm 0.03$.
The expected position of the modes according to this solution is depicted 
with dashed lines in Figure~4.
Figure~5 shows a  close-up of the power spectrum with the identified 
frequencies according with the modified extraction. 
\subsection{Oscillation amplitudes}
The strongest peaks in the amplitude spectrum of \muher{} (square root of
power) reach about 0.9\,\ms.  However, these are likely to have been
strengthened significantly by constructive interference with noise peaks.
As stressed by \citet{K+B95} the effects of the noise must
be taken into account when estimating the amplitude of the underlying
signal.  In order to estimate the  average amplitude of the modes we  used the 
result of \citet{K+B95} (see also \citealt{KBB2005} ), where the effect of the noise
is estimated. By considering  the five strongest peaks 
we found a mean amplitude of the $p$-mode peak power, 
for the modes with $l=0,1$ in the frequency interval $0.8-1.6\,\rm mHz$, 
of $(63.14\pm 3.0)\,\rm cm\,s^{-1}$.  The observed amplitudes  
are thus in good agreement with the theoretical amplitudes obtained 
by \citet{K+B95}   for which $v_{\rm osc} 
=  (L/L_\odot)/(M/M_\odot) \; (23.4\pm 1.4) \; \rm cm \;s^{-1} = (69.50\pm 4.1) \; \rm cm\;s^{-1}$ 
(see also the discussion by \citealt{Hou2006}).  
We also note that the frequency of excess of power 
(1.2\,mHz) is consistent with the value expected
from scaling the acoustic cutoff from the solar case. 

\section{Conclusion}
Our observations of \muher{} show an evident excess of power in the PS, clearly
disentangled from the low frequency increase, and with a position and amplitude that are
in agreement with expectations.  Although hampered by the single-site
window, the  comb analysis and the echelle diagram show clear evidence for regularity in the
peaks at the spacing expected from asymptotic theory.  
Moreover, our results provide valuable confirmation 
that oscillations in solar-like stars really do have the amplitudes that 
scales as $L/M$  by extrapolating from the Sun.  
We hope that in the near future a multi-site observing campaign 
will allow us to explore further the oscillation spectrum of \muher. 

\acknowledgments
We would like to thank Tim Bedding and Antonio Frasca 
for important suggestions and encouragements.
This work was supported financially by the INAF grant ``PRIN - 2006".


%
%

\begin{table*}
\begin{center}
\caption{\label{tab.freqs}Oscillation frequencies in \muher{} 
for the modified extraction (left) and standard extraction (right)}
\begin{tabular}{rcl|lccl}
\tableline
\noalign{\smallskip}
%
%
\tableline
(\muHz)    &  S/N & \sc Mode ID & (\muHz)   &  S/N & \sc Mode ID \\ 
\noalign{\smallskip}
\tableline
\noalign{\smallskip}
 985.5(0.2)     &  5.3 & $l=  0$, $n=16$ & 985.6(0.2) &   4.9 & $l=  0$, $n=16$ \\
 1100.4(0.3)    &  3.8 & $l= 0$, $n=18$ &  & &      \\
 1211.2(0.1)    &  9.1 & $l= 0$, $n=20$ & 1211.2(0.1) &  9.1    & $l=  0$, $n=20$                 \\
 1268.7(0.3)    &  3.1 & $l= 0$, $n=21$ &        &         &                \\
 1323.2(0.3)    &  2.9 & $l= 0$, $n=22$ & 1323.4(0.3) &  3.5 & $l=  0$, $n=22$ \\
 1437.6(0.3)    &  3.5 & $l= 0$, $n=24$ & 1437.8(0.2) =1426.1+11.6  &   5.3 & $l=  0$, $n=24$ \\
                &      &                & 1608.2(0.3) =1596.6+11.6  &   3.3 & $l=  0$, $n=27$ \\
 1125.1(0.2)  &  4.1 & $l= 1$, $n=18$ & 1124.8(0.1) =1136.4-11.6 &   6.9 & $l=  1$, $n=18$ \\
 1180.0(0.2)  &  4.2 & $l= 1$, $n=19$ & 1180.0(0.2) =1192.4-11.6  & 5.0  & $l=1$, $n=19$   \\
 1238.9(0.3)  &  3.8 & $l= 1$, $n=20$ &        &    &                           \\
 1296.4(0.3)  &  3.0 & $l= 1$, $n=21$ &        &    &                            \\
 1351.9(0.2) &  4.4 & $l= 1$, $n=22$ & 1352.1(0.2) &   4.7 & $l=  1$, $n=22$ \\
 1034.0(0.2) & 3.2 & $l=2$, $n=16$  & 1034.3(0.3) =1045.9-11.6 &   3.3 & $l=  2$, $n=16$ \\
         &      &                & 1205.6(0.3) &   3.2 & $l=  2$, $n=19$ \\
         &      &                & 1375.4(0.3)=1387.0-11.6 &   3.4 & $l=  2$, $n=22$ \\
 947.6(0.2)   &  4.0 & $l= 3$, $n=14$ & 947.5(0.3)  &   3.9 & $l=  3$, $n=14$ \\
 1061.0(0.2)  &  4.0 & $l= 3$, $n=16$ & 1061.2(0.3) &   3.9 & $l=  3$, $n=16$ \\
 1173.0(0.3)  & 3.2  & $l=3$,  $n=18$ & 1172.9(0.3) &   3.1 & $l=  3$, $n=18$ \\
         &      &                & 1285.1(0.2) &   5.1 & $l=  3$, $n=20$ \\
         &      &                & 1569.2(0.3) &   3.5 & $l=  3$, $n=25$ \\
\noalign{\smallskip}
\tableline
\end{tabular}
\end{center}
\end{table*}

%
%



\begin{figure*}
\epsscale{1} 
\plotone{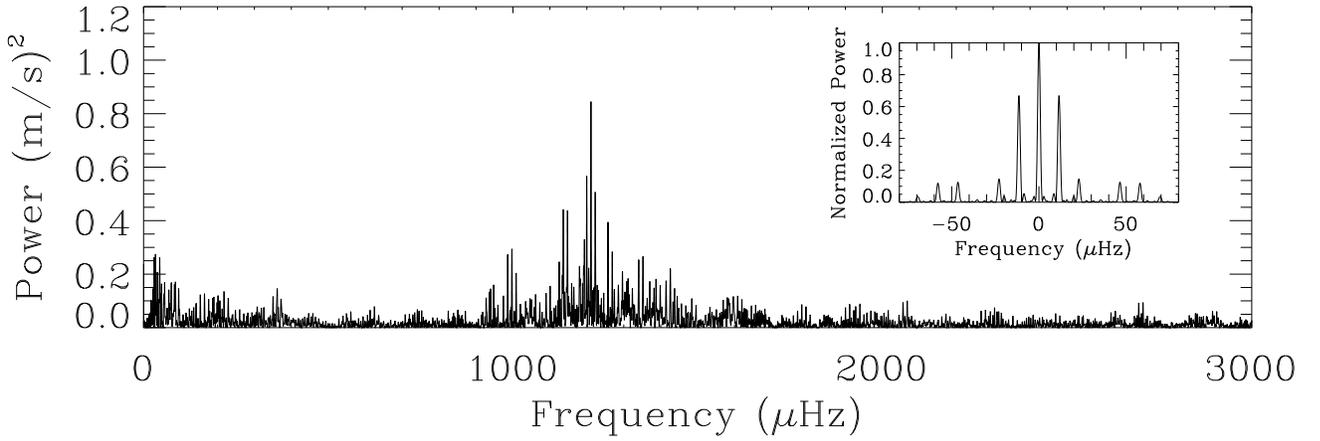}
\caption[]{\label{fig2} Power spectrum of the SARG velocity
measurement of \muher{} with the spectral window (in the inset).  }
\end{figure*}

\begin{figure*}
\epsscale{.4} 
\plotone{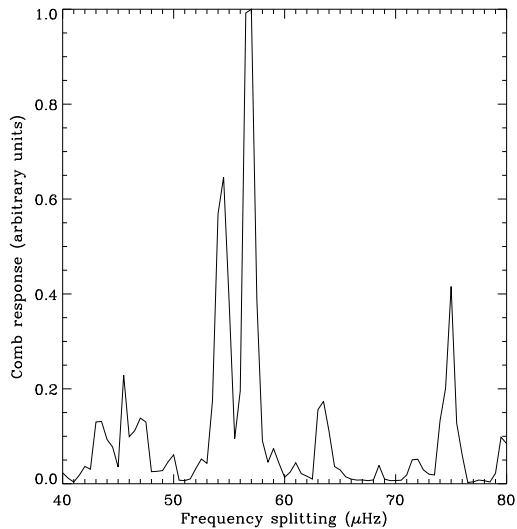}
\caption[]{\label{fig3} CR function of the power spectrum of the \muher{} data.
 }
\end{figure*}

\begin{figure*}
\epsscale{.9} 
\plotone{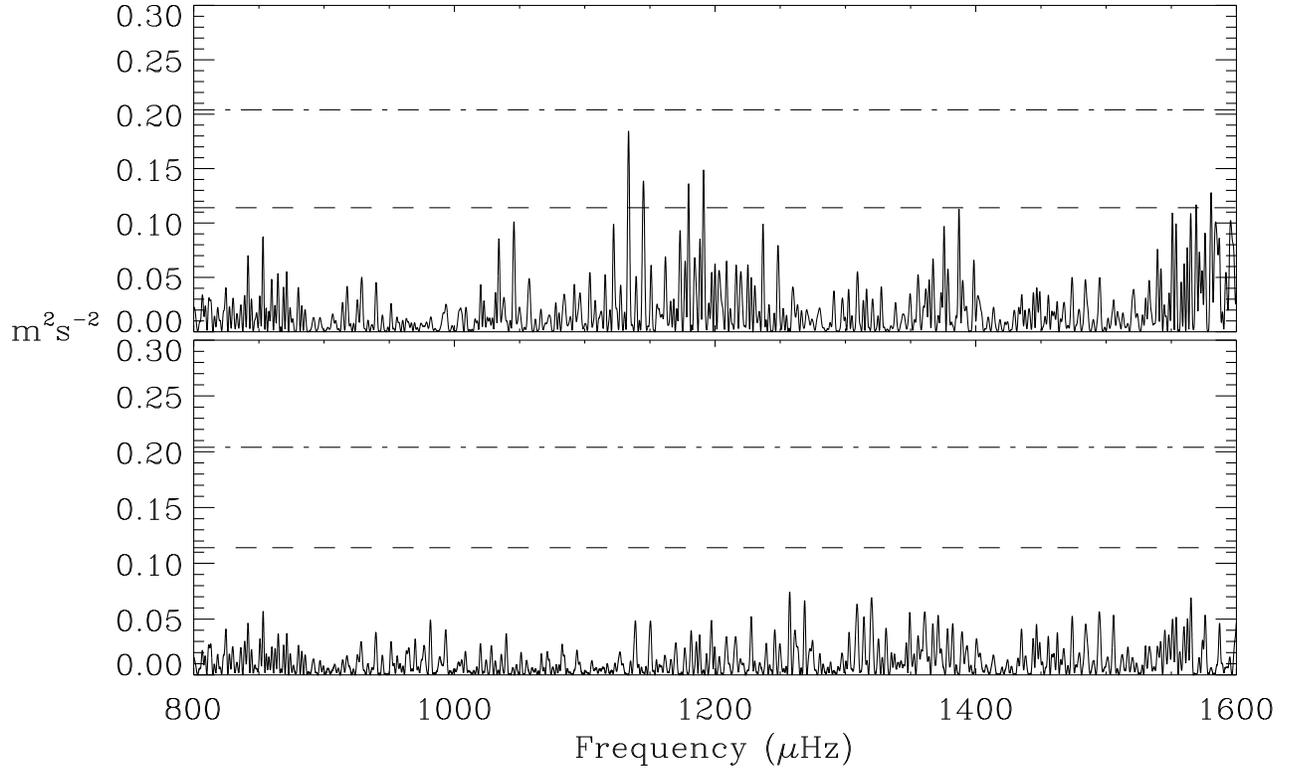}
\caption[]{\label{fig4}
The power spectrum of the residuals obtained with the modified extraction
(upper panel) and with the standard extraction (lower panel). 
The dot-dashed and the dashed lines indicate $4\,\sigma$
and $3\,\sigma$ respectively.   
 }
\end{figure*}

\begin{figure*}
\epsscale{.8} 
\plotone{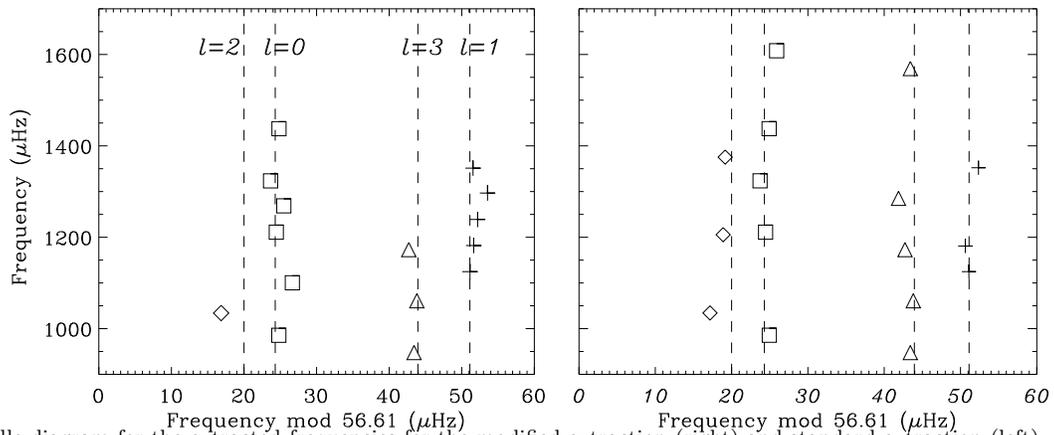}
\caption[]{\label{fig5} Echelle diagram for the extracted frequencies for the modified 
extraction (right) and standard extraction (left). Squares, crosses, rhomboes, triangles
are for $\ell=0,1,2,3$ respectively. The dashed vertical lines represent the position 
of the $\ell=0,1,2,3$ modes according to our fit of the asymptotic relation.
 }
\end{figure*}

\begin{figure*}
\epsscale{.9} 
\plotone{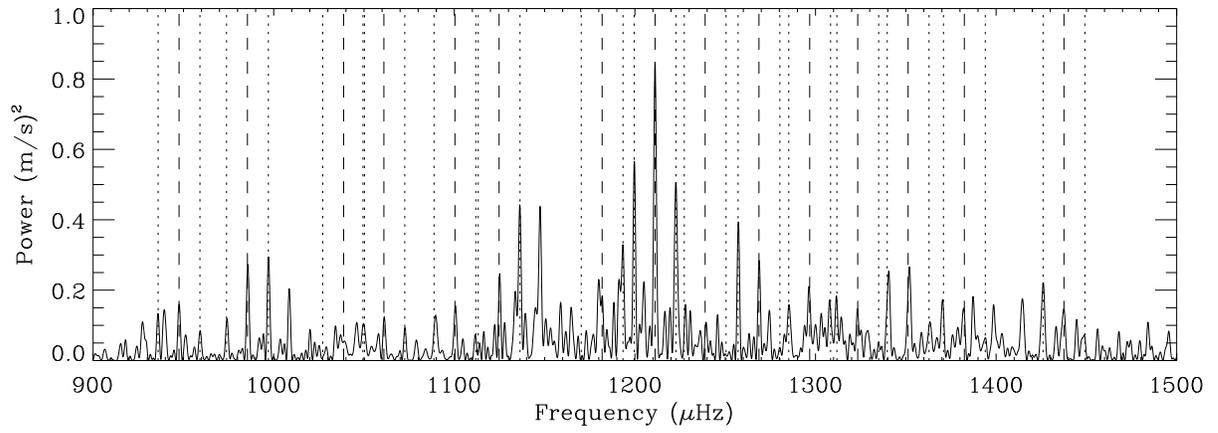}
\caption[]{\label{fig6} Close-up of the spectrum of \muher{} with the 
identified frequencies.  Dotted lines are the day-night aliases.
 }
\end{figure*}

\end{document}